\documentclass{elsart}
\usepackage{deluxetable}

\usepackage{graphicx}
\usepackage{amsmath}
\usepackage{amsfonts}
\usepackage{amssymb}
\usepackage{setspace}
\usepackage{amssymb}
\begin{document}
\begin{frontmatter}
\journal{Astroparticle Physics}


\newcommand{\bremm}{Bremsstrahlung}
\newcommand{\addressoficrr}[1]{$^{a}$ #1 }
\newcommand{\addressofncen}[1]{$^{b}$ #1 }
\newcommand{\addressofbu}[1]{$^{c}$ #1 }
\newcommand{\addressofbnl}[1]{$^{d}$ #1 }
\newcommand{\addressofuci}[1]{$^{e}$ #1 }
\newcommand{\addressofcsu}[1]{$^{f}$ #1 }
\newcommand{\addressofcnu}[1]{$^{g}$ #1 }
\newcommand{\addressofduke}[1]{$^{h}$ #1 }
\newcommand{\addressofgifu}[1]{$^{i}$ #1 }
\newcommand{\addressofuh}[1]{$^{j}$ #1 }
\newcommand{\addressofui}[1]{$^{k}$ #1 }
\newcommand{\addressofkek}[1]{$^{l}$ #1 }
\newcommand{\addressofkobe}[1]{$^{m}$ #1 }
\newcommand{\addressofkyoto}[1]{$^{n}$ #1 }
\newcommand{\addressoflanl}[1]{$^{o}$ #1 }
\newcommand{\addressoflsu}[1]{$^{p}$ #1 }
\newcommand{\addressofmit}[1]{$^{r}$ #1 }
\newcommand{\addressofduluth}[1]{$^{s}$ #1 }
\newcommand{\addressofmiyagi}[1]{$^{t}$ #1 }
\newcommand{\addressofsuny}[1]{$^{u}$ #1 }
\newcommand{\addressofnagoya}[1]{$^{v}$ #1 }
\newcommand{\addressofniigata}[1]{$^{w}$ #1 }
\newcommand{\addressofokayama}[1]{$^{x}$ #1 }
\newcommand{\addressofosaka}[1]{$^{y}$ #1 }
\newcommand{\addressofseoul}[1]{$^{z}$ #1 }
\newcommand{\addressofshizuokaseika}[1]{$^{aa}$ #1 }
\newcommand{\addressofshizuoka}[1]{$^{bb}$ #1 }
\newcommand{\addressofskku}[1]{$^{cc}$ #1 }
\newcommand{\addressoftohoku}[1]{$^{dd}$ #1 }
\newcommand{\addressoftokyo}[1]{$^{ee}$ #1 }
\newcommand{\addressoftokai}[1]{$^{ff}$ #1 }
\newcommand{\addressoftit}[1]{$^{gg}$ #1 }
\newcommand{\addressofwarsaw}[1]{$^{hh}$ #1 }
\newcommand{\addressofuw}[1]{$^{ii}$ #1 }
\title{Study of TeV Neutrinos with Upward Showering Muons in Super-Kamiokande}

\author[bu]{S.~Desai\thanksref{psu}\corauthref{cor}}
%
\author[icrr]{K.~Abe}
\author[icrr]{Y.~Hayato}
\author[icrr]{K.~Iida}
\author[icrr]{K.~Ishihara}
\author[icrr]{J.~Kameda}
\author[icrr]{Y.~Koshio}
\author[icrr]{A.~Minamino}
\author[icrr]{C.~Mitsuda}
\author[icrr]{M.~Miura}
% ...
\address[icrr]{Kamioka Observatory, Institute for Cosmic Ray Research, University of Tokyo, Kamioka, Gifu, 506-1205, Japan}
% ...
\address[bu]{Department of Physics, Boston University, Boston, MA 02215, USA}
% ...

\collaboration{The Super-Kamiokande Collaboration}

\thanks[psu]{Present address: Department of Physics, Pennsylvania State University, University Park, PA 16802, USA}
\corauth[cor]{Corresponding author. Email: \texttt{shantanu@neutrino.bu.edu}}
\date{\today}

\begin{abstract}
A subset of neutrino-induced upward through-going muons in the
Super-Kamiokande detector consists of high energy muons which lose energy
through radiative processes such as  bremsstrahlung, $e^{+} e^{-}$ pair
production and photonuclear interactions.  These ``upward showering muons''
comprise an event sample whose mean parent neutrino energy is approximately
1 TeV. We show that the zenith angle distribution of upward showering muons
is consistent with negligible distortion due to neutrino oscillations, as expected of
such a high-energy neutrino sample. We present astronomical searches using 
these high energy events, such as those from WIMP annihilations in the Sun, Earth and
Galactic Center, some suspected point sources, as well as searches for
diffuse flux from the interstellar medium.
\end{abstract}
\end{frontmatter}

\section{INTRODUCTION}

In order to select neutrino events with the highest energies, we 
consider muon neutrino
interactions in the rock around the detector, because the
effective target volume is very much increased~\cite{Grillo}.  To separate
neutrino-induced muons from cosmic ray muons, we select only
upward-going muons, since the background from downward going cosmic ray muons
overwhelms any neutrino-induced muons from above. Muons penetrating the
detector have energies of at least several GeV and point along the neutrino
direction within a few degrees, allowing astrophysical studies.  Neutrinos
originating from cosmic point sources are expected to have harder energy
spectra than the background of atmospheric neutrinos.  Some of these high 
energy
neutrino-induced muons undergo radiative energy loss. We identify these 
muons  as ``showering muons''. 
High energy muons are
correlated with high energy neutrinos, which allow us to statistically select
an extremely high energy parent neutrino sample from the Super-Kamiokande data. Our aim
is to extract this sample for physics and astronomy studies.

The Super-Kamiokande (Super-K) experiment~\cite{sknim} has previously
analyzed two topologically distinct categories of upward muons caused by
neutrino interactions in the rock below the detector: muons which exit the
detector (called ``through-going'') and those which stop inside the detector
(called ``stopping'')~\cite{skstop}. The parent neutrino energy of upward
stopping and through-going muons for atmospheric neutrinos is peaked at $\simeq$ 10 GeV and $\simeq$
100 GeV respectively.  In 1646 days of data, Super-Kamiokande detected 1856
upward through-going and 458 upward stopping muons.  Oscillation results
using only upward through-going muons and stopping muons have been reported in
Refs.~\cite{skstop} and ~\cite{upthru}. Both upward muon samples have been
combined along with the contained events to do the most precise atmospheric neutrino
oscillation studies~\cite{fullpaper,3flavor}. Moreover, astrophysical 
searches for
annihilation signatures of WIMPs~\cite{skwimp}, point and diffuse
sources~\cite{sknuastro}, and neutrinos from GRBs~\cite{skgrb} have been done
with upward muons.
 
In this paper we bifurcate the upward through-going muon sample into
upward showering and upward non-showering events.  This
classification is done on the basis of the physical mechanism for muon energy
loss above 1 TeV. The showering sample is then analyzed to see what these
highest-energy events might reveal.

\section{MUON ENERGY LOSS}

At low energies, from 100 MeV to 100 GeV, muon energy loss is dominated by
atomic ionization as described by Bethe-Bloch; the minimum ionization energy
loss in water is 1.99 MeV/cm~\cite{pdg}. At higher energies, radiative
processes such as $e^{+}e^{-}$ pair production, bremsstrahlung,~and
photonuclear interactions become important, resulting in catastrophic energy
losses with large fluctuations. In water, the critical energy where
ionization energy loss is equal to the average radiative energy loss is 
1.03~TeV~\cite{Groom}. The relevant processes have recently been reviewed in
Refs.~\cite{Lipari,Seckel}.

\begin{figure}
\begin{center}
    \includegraphics[height=17.0pc]{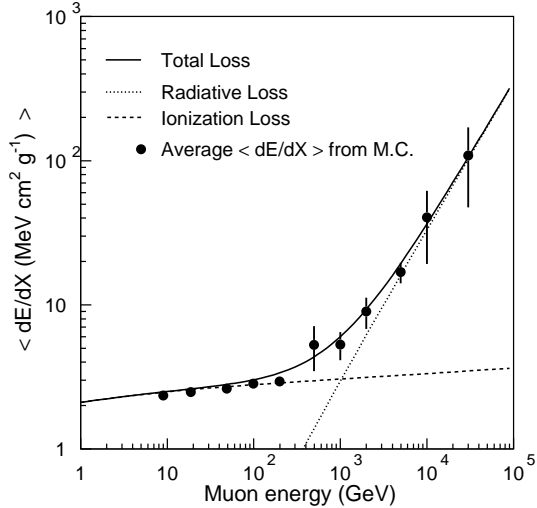}
\end{center}
  \vspace{-0.5pc}
  \caption{\label{neutenergyloss} Average muon energy loss/tracklength
for about 100 simulated muons at various energies from 10 GeV to 30 TeV.
The theoretical curves for energy loss due to ionization and other
radiative processes have been obtained from Ref.~\cite{Groom}. }
\end{figure}

To develop an algorithm for showering/non-showering separation and test its
efficiency, we used a {\tt GEANT~3} based  detector simulation to simulate muons at several energies between 
10~GeV and 30~TeV. We checked this Monte Carlo sample to make sure that the muon energy loss in
our detector simulation  agrees with calculated values\cite{Groom}. The average energy loss/tracklength at each
muon energy is shown in Fig.~\ref{neutenergyloss}, in good agreement with the
calculation.

All details of the  muon energy loss calculations for different 
processes in {\tt GEANT~3} are discussed in Ref.~\cite{lohmann}. The
cross-section for bremsstrahlung is  obtained from Ref.~\cite{petrukhin}.
The energy loss for direct electron-positron pair production is
obtained from Ref.~\cite{kokoulin}. Cross-sections for  photonuclear
interactions are obtained from Ref.~\cite{bugaev}.

\section{DATA AND MONTE CARLO USED}

For this paper, we used all upward through-going muons from the full
Super-K-I dataset from May 1996 to July 2001 spanning 1645.9 days
of data.  We require that  reconstructed tracks have a minimum length of 7~m,
which reduces the contamination  from photo-produced upward pions to 
0.01\%~\cite{macro98,sdthesis}. 
This path-length cut of 7~m corresponds to a muon energy threshold of 1.6 GeV. As discussed in
Ref.~\cite{fullpaper}, all upward muon events require between 8,000 
and
1,750,000 photoelectrons. A specialized analysis looking for upward muons in
events with greater than 1,750,000 photoelectrons has been reported
elsewhere~\cite{skuheupmu}.

We applied the showering algorithm to simulated upward
through-going muons from the standard Super-Kamiokande atmospheric neutrino
Monte Carlo sample~\cite{fullpaper}. The input atmospheric neutrino flux used
for this Monte Carlo sample was from Ref.~\cite{Honda} for neutrino energies
up to 1 TeV. At 1 TeV, the calculated flux in Ref.~\cite{Volkova} was
rescaled to that in Ref.~\cite{Honda}, and used for neutrino energies up to
100 TeV.  The uncertainty in the absolute atmospheric neutrino flux is 
about
20\% below 1 TeV and and ranges from 25\%~\cite{Honda07} to 40\%~\cite{Gaisser06,Barr06} around 1 TeV.
The atmospheric neutrino flux in Ref.~\cite{Honda07} is about 15-20\%
greater at 1 TeV as compared to Ref.~\cite{Honda}. Neutrino interactions with the
rock outside the detector (assumed to be ``standard rock'') as well as in the
water were simulated using the {\tt NEUT}~\cite{NEUT} interaction package.
Neutrino interactions in the rock were simulated up to 4~km from the center
of the detector.  The total livetime simulated was equal to 100 years.

\section{LIGHT PRODUCTION MODEL}

Below the critical energy, an ionizing muon produces a constant amount of
Cherenkov light per unit tracklength.  However, a muon which undergoes
radiative energy loss produces high-energy photons which create 
electron-positron pairs, thus increasing the total Cherenkov
light in the detector. If we can calibrate the total Cherenkov light
emitted by a normally ionizing muon (after accounting for the various sources
of light attenuation in Super-K) then any electromagnetic shower associated
with the muon will emit excess light over this amount. Thus we need to
calibrate the total ``luminosity'' of ionizing muons with energy well below
the critical energy of 1 TeV.

Using the muon entry point and direction as inputs, and accounting for the effective water
attenuation length ($L_{att}$) and geometrical acceptance corrections, we apply various
corrections to the raw PMT photoelectrons.  The corrected number of
photoelectrons of each PMT in the Cherenkov cone is:
\begin{equation}
\label{chargecorrection}
q_{corr} \mbox{(corr.~pe)} = K \frac{q_{raw} d_{w} e^{\left(\frac{d_{w}}{L_{att}}\right)}}{F(\theta)},
\end{equation}
where $q_{raw}$ is the raw number of photoelectrons detected by each PMT;
$d_{w}$ is the distance traveled by the photons from the point along the muon
track where the photon is emitted to the PMT which detects it; $F(\theta)$
accounts for the PMT angular acceptance and shadowing;
and $K$ is a normalization constant (=~1/2500 ${\rm cm^{-1}}$) which makes 
the
corrected photoelectrons the same order of magnitude as the original raw
photoelectrons. With these corrections to the charge of each PMT, the units
of $q_{corr}$ and other terms obtained from $q_{corr}$ are corrected
photoelectrons.

\begin{figure}
\begin{center}
    \includegraphics[height=17.0pc]{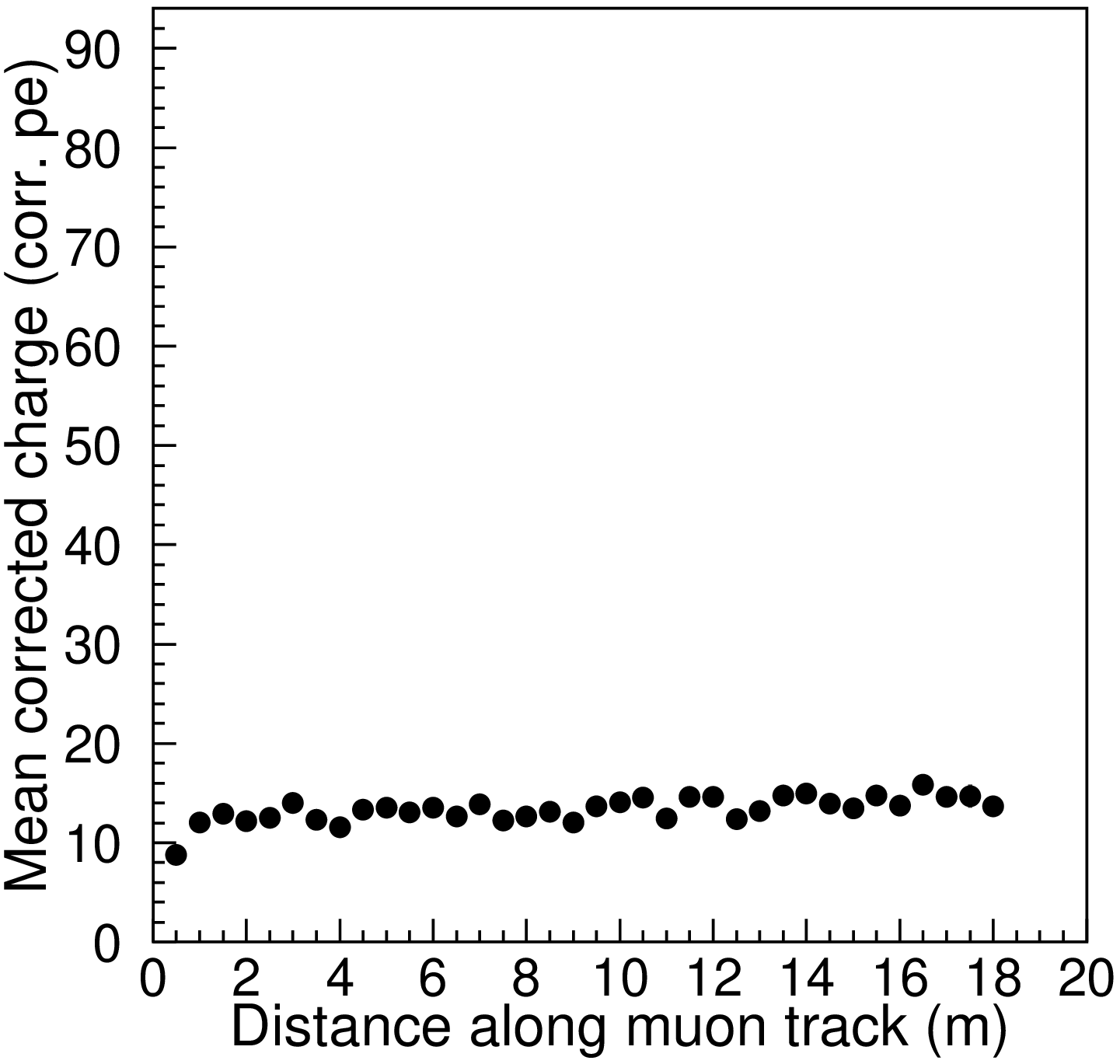}
   \includegraphics[height=17.0pc]{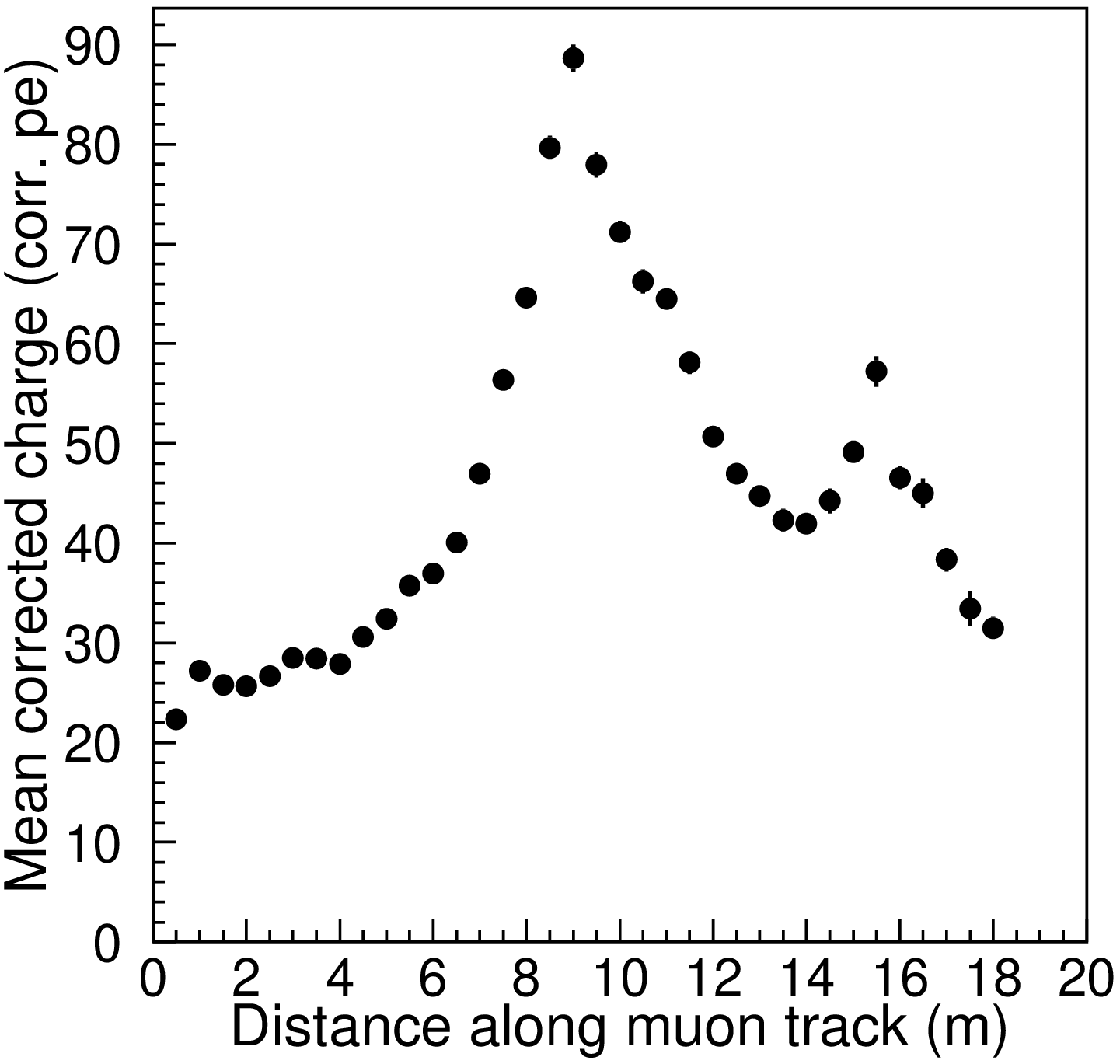}
\end{center}
  \vspace{-0.5pc}
  \caption{\label{ionizing} The corrected photoelectrons
    for an example simulated ionizing muon with energy 20 GeV (top) and a
    showering muon with energy 10 TeV (bottom) simulated with the same entry
    point and direction.}
\end{figure}

We then calculate the average charge in small tracklength intervals (50~cm) along
the muon track:
\begin{equation}
 Q_{corr}^i  = \frac{1}{N_{pmt}} \sum_{k=1}^{N_{pmt}}{q_{corr}}, 
\label{Qcorr}
\end{equation}
where $N_{pmt}$ is the number of
PMTs whose projected distance along the muon track is within the 50~cm
path-length interval.  
The statistical uncertainty in $Q_{corr}^i$ is given by:
\begin{equation}
\sigma_{Q_{corr}^i}^2   = \frac{1}{N_{pmt}^2} \sum_{k=1}^{N_{pmt}} \frac{\left(q_{corr}\right)^2}{q_{raw}},
\label{sigmaqcorr}
\end{equation}
assuming that the error in raw PMT charge arises from the square-root of the
number of photoelectrons. We then plot $Q_{corr}^i$ along the muon track.  An
example plot for a typical ionizing Monte Carlo muon event can be seen in the
top panel of Fig.~\ref{ionizing}.
We also calculate the average of $Q_{corr}^i$ along the muon track as follows:
\begin{equation}
 \langle Q_{corr} \rangle = \frac{\sum \limits_{i=4}^{N-3}
   \left(Q_{corr}^i/\sigma_{Q_{corr}^i}^2\right)}{\sum \limits_{i=4}^{N-3}1/\sigma_{Q_{corr}^i}^2},
\label{qcorravg}
\end{equation}
where $N$ indicates the number of 50~cm bins in which the muon track is 
sub-divided. In Eqn.~\ref{qcorravg}, the sum goes from 4 to $N-3$, since the first and last 1.5~m
from the muon track are excluded. This is because corrections
for PMT acceptance and shadowing are not accurately modeled near the wall.

In addition, because of the effects of light scattering into the
Cherenkov cone, longer path-length muons have greater corrected mean
charge compared to short path-length muons. This is accounted for by
using a path-length correction to the average charge of an ionizing
muon.  The corrected charge distribution of a typical ionizing muon is
compared to that of a  showering muon with the same entry point and
direction in the bottom panel of Fig.~\ref{ionizing}. Our aim is to
construct a function which quantitatively distinguishes between these
based on shape and total collected photoelectrons.

\section{SELECTION ALGORITHM FOR SHOWERING MUONS}
\label{showeringalgo}
 We define a showering $\chi^2$  as follows: 
\begin{equation}
\label{showeringchisquare}
\chi_{showering}^2 = \frac{1}{N-6}
\sum_{i=4}^{N-3}\left\{\frac{\left[Q_{corr}^i-\langle Q_{corr} \rangle 
\right]}{\sigma_{Q_{corr}^i}}\right\}^2 , \end{equation}
where  $N$ is same as in Eqn.~\ref{qcorravg}; $\langle Q_{corr} \rangle$ 
is defined in Eqn.~\ref{qcorravg}; $\sigma_{Q_{corr}^i}$ is the 
statistical error in $Q_{corr}^i$ and defined in Eqn.~\ref{sigmaqcorr}. 
Equation~\ref{showeringchisquare} measures deviations in the 
histogram from a flat distribution.  

\begin{figure}
\begin{center}
\includegraphics[height=17.0pc]{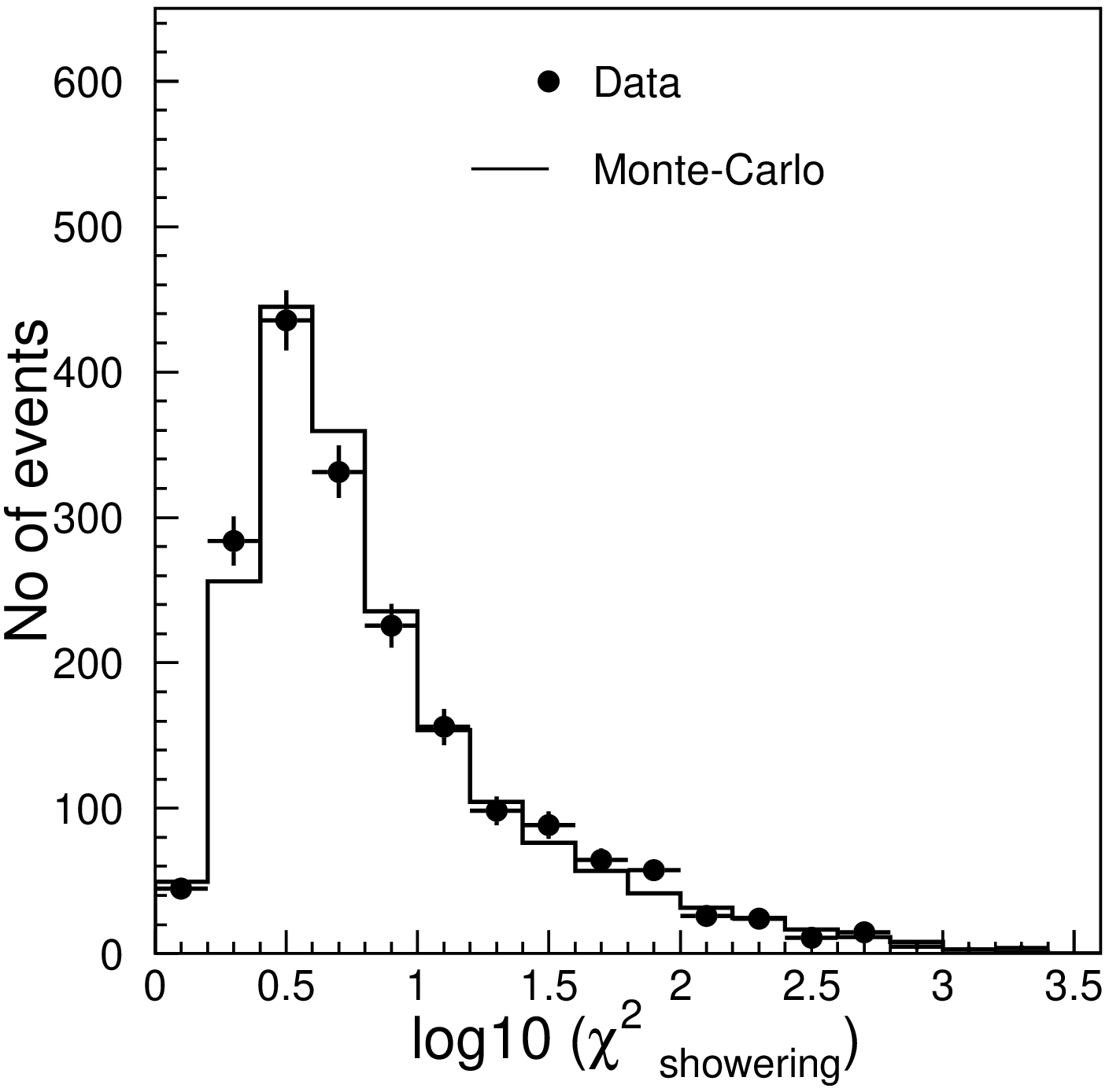}
\includegraphics[height=17.0pc]{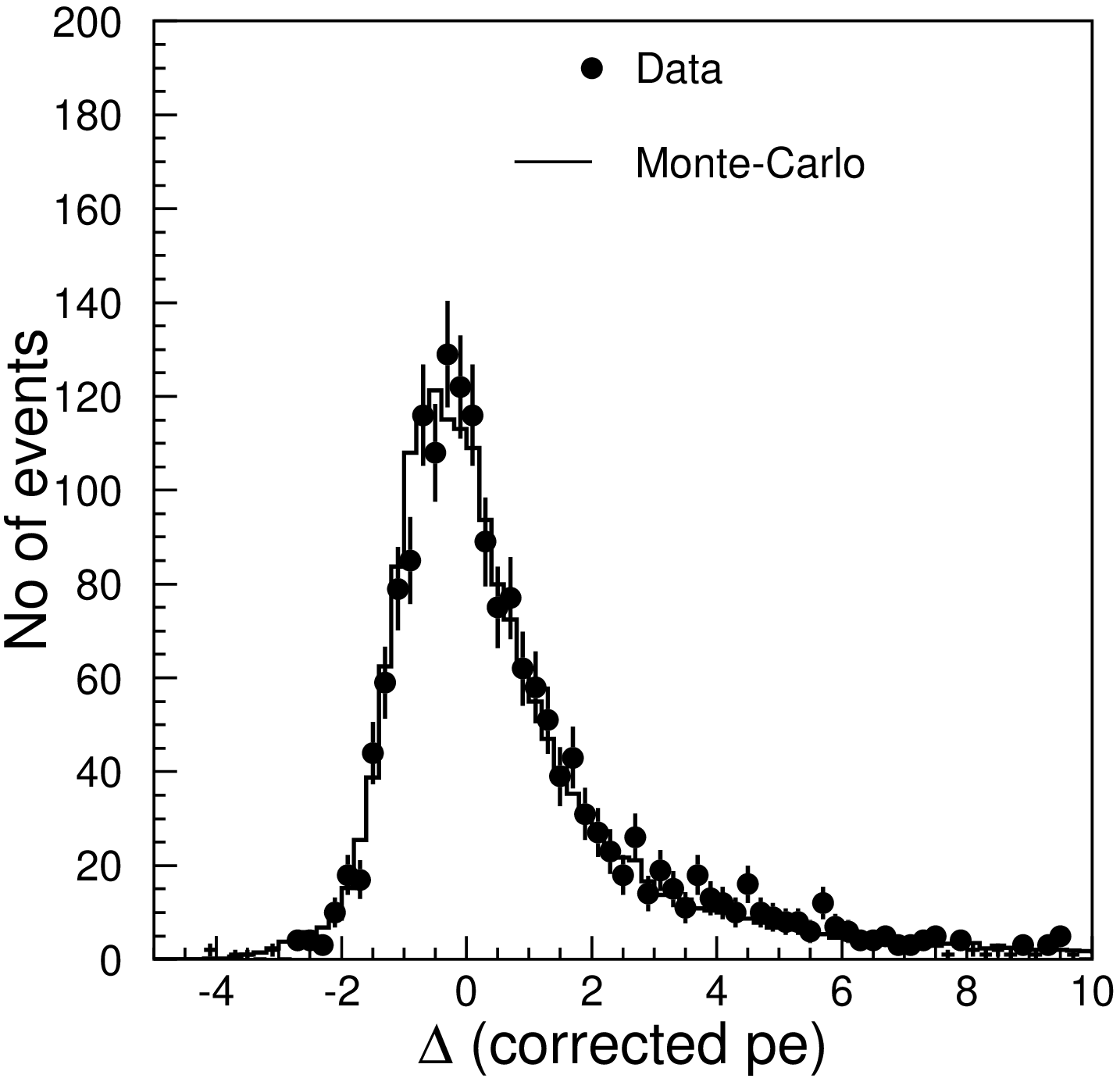}
\end{center}
\vspace{-0.5pc}
\caption{\label{showeringvar} Comparison of showering variables
as described in Eqn.~\ref{showeringchisquare} and 
~\ref{delta} for Monte Carlo and 1645.9 days upward 
through-going muon data. The solid line is normalized to total number of events.}
\end{figure}

About 0.5\% of events in the Monte Carlo consist of  through-going muons with bad fits or 
stopping muons which are misclassified as through-going muons. These events show large values of $\chi^2$ 
even
though the total number of photoelectrons in such events is very small. To eliminate
these cases, we define another variable to distinguish between a showering
and non-showering muon:
\begin{equation}
\label{delta}
\Delta =  [\langle Q_{corr} \rangle - Q_{exp}(l)] , 
\end{equation}
where $ \langle Q_{corr} \rangle$ is defined in Eqn.~\ref{qcorravg}
and $Q_{exp}(l)$ is a path-length dependent estimate of the
expected charge of a normally ionizing muon.  The magnitude of
$\Delta$ indicates how much more Cherenkov light is present relative
to a normally ionizing muon. The comparison of these variables for
data and 100 year atmospheric neutrino Monte Carlo is shown in
Fig.~\ref{showeringvar}. The distributions of these
variables for data and Monte Carlo is in good agreement.

\begin{figure}
\begin{center}
    \includegraphics[height=17.0pc]{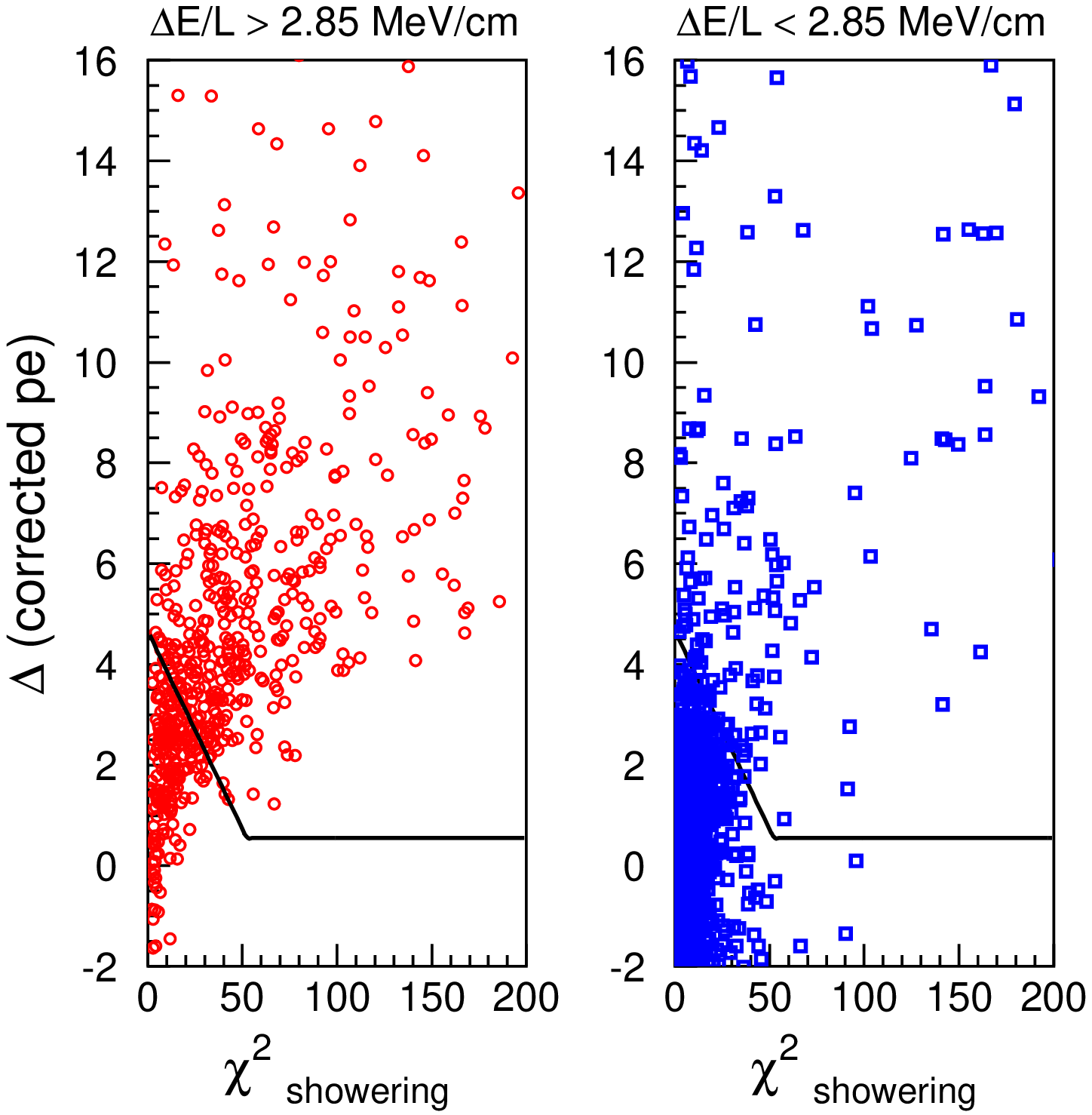}
\end{center}
  \vspace{-0.5pc}
  \caption{\label{showerscatter} Distribution of $\Delta$ vs $\chi^2_{showering}$ 
for events from  a   10 year Monte Carlo sample (which is a subset of the
100 year atmospheric neutrino Monte Carlo) with true $\Delta E/L >$ 2.85 MeV/cm (left panel),
i.e. ``true showering events,'' and true  $\Delta E/L <$ 2.85 MeV/cm (right panel), i.e. ``true 
non-showering events.'' The solid line indicates the cut used to
separate showering from non-showering muons.} 
\end{figure}

We applied this algorithm to the 37287 upward through-going muon
events in the 100 year neutrino Monte Carlo~\cite{fullpaper}.  The scatter
plots of the distribution of $\chi^2_{showering}$ and $\Delta$ 
are shown in Fig.~\ref{showerscatter} for a 10 year
subset of the 100 year Monte Carlo. They are presented separately for samples of ``true'' showering and 
non-showering muons, which are defined based on the true muon energy loss 
per unit tracklength $\Delta E/L$ (where $\Delta E/ L $ is the true
muon energy loss in the inner detector for a muon of path-length $L$), 
with the separation  value equal to 2.85 MeV/cm.


The separation between showering
and non-showering events is decided by the line shown in Figure.~\ref{showerscatter}. 
With this cut we found a total of 5747 upward showering events in the
Monte Carlo.
We then calculate the efficiency, defining true showering muons as those
with $\Delta E/ L >$ 2.85~MeV/cm. 
This cut selects about 70\% of such events, while the
misidentification of events with true $\Delta E/ L < 2.85$~MeV/cm  is
about 5\%. These cuts provide reasonable purity and background
contamination in forming a sample for astrophysical studies and
neutrino oscillation.

\begin{figure}
\begin{center}
    \includegraphics[height=17.0pc]{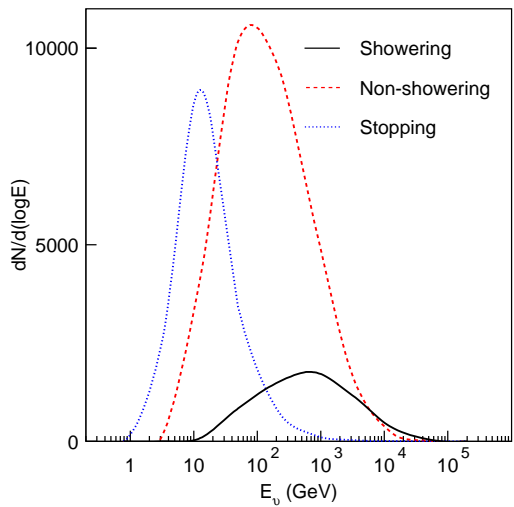}
\end{center}
  \vspace{-0.5pc}
  \caption{\label{energyspectrum} Neutrino energy spectra of all 3
    categories of upward muons in Super-Kamiokande in 100 yrs of atmospheric
neutrino Monte Carlo.} 
\end{figure}

The parent neutrino energy spectra of all upward stopping, non-showering
through-going, and showering through-going muons are shown in
Fig.~\ref{energyspectrum}. The mean parent neutrino energy of
upward showering muon events is peaked at 1 TeV. Thus the showering dataset
constitutes the highest energy neutrinos seen in Super-K. The total number
of upward showering muons, is approximately one-fifth the total
number of upward through-going muons.

The estimated muon angular resolution of upward
showering muon events is about $1.4^{\circ}$, where angular resolution is defined as the 
average angular separation between true muon direction and reconstructed muon direction. 
The corresponding angular resolution for upward stopping and through-going muons is 
about $2.4^{\circ}$ and $1.3^{\circ}$ respectively.
For higher energy muons, the additional photons generated through radiative processes make it harder to 
reconstruct the muon track direction. Hence the angular resolution is 
slightly worse for showering muons compared to the non-showering muons.

\section{BACKGROUND SUBTRACTION}
\begin{figure}
\begin{center}
    \includegraphics[height=17.0pc]{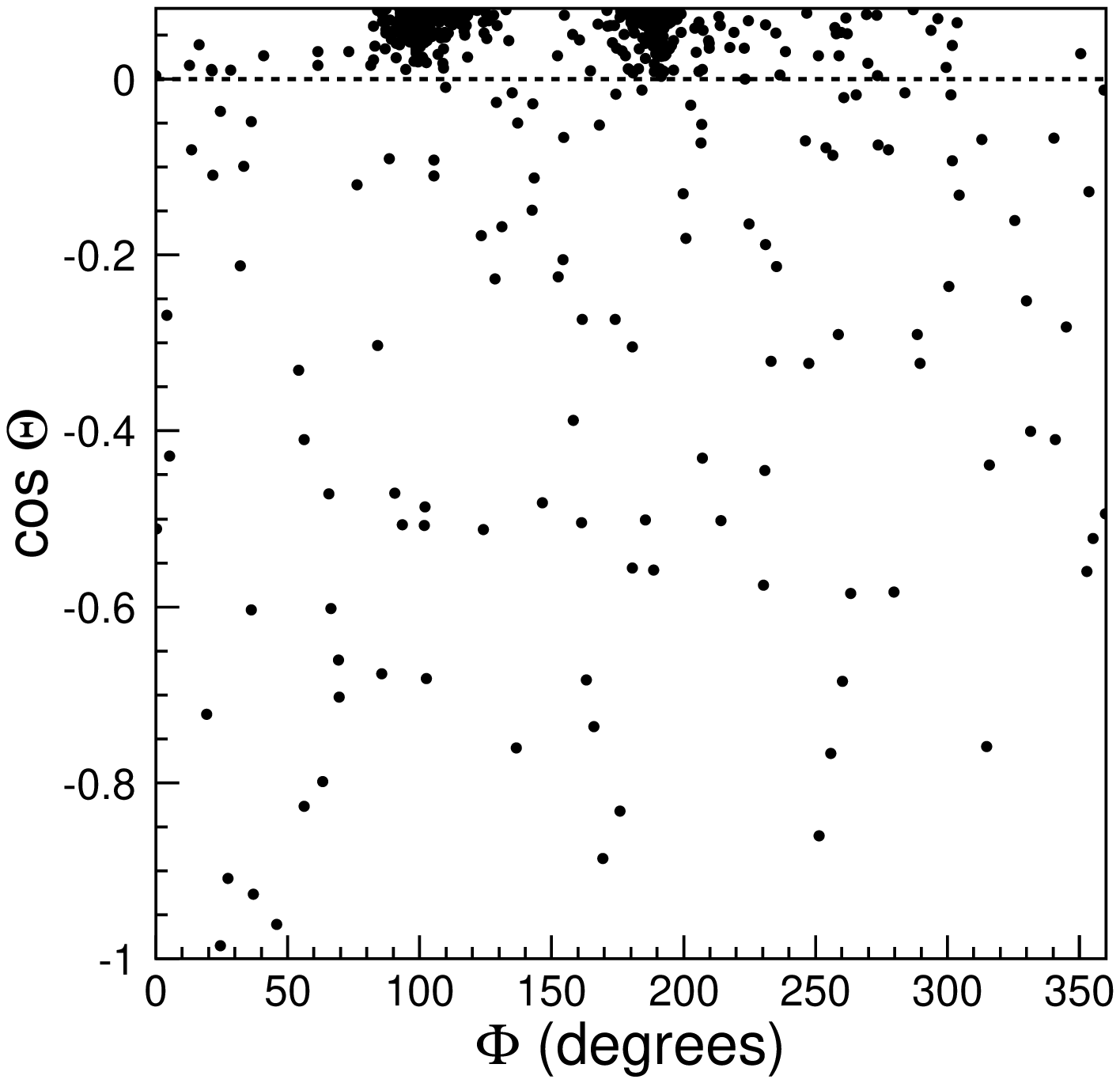}
 \includegraphics[height=17.0pc]{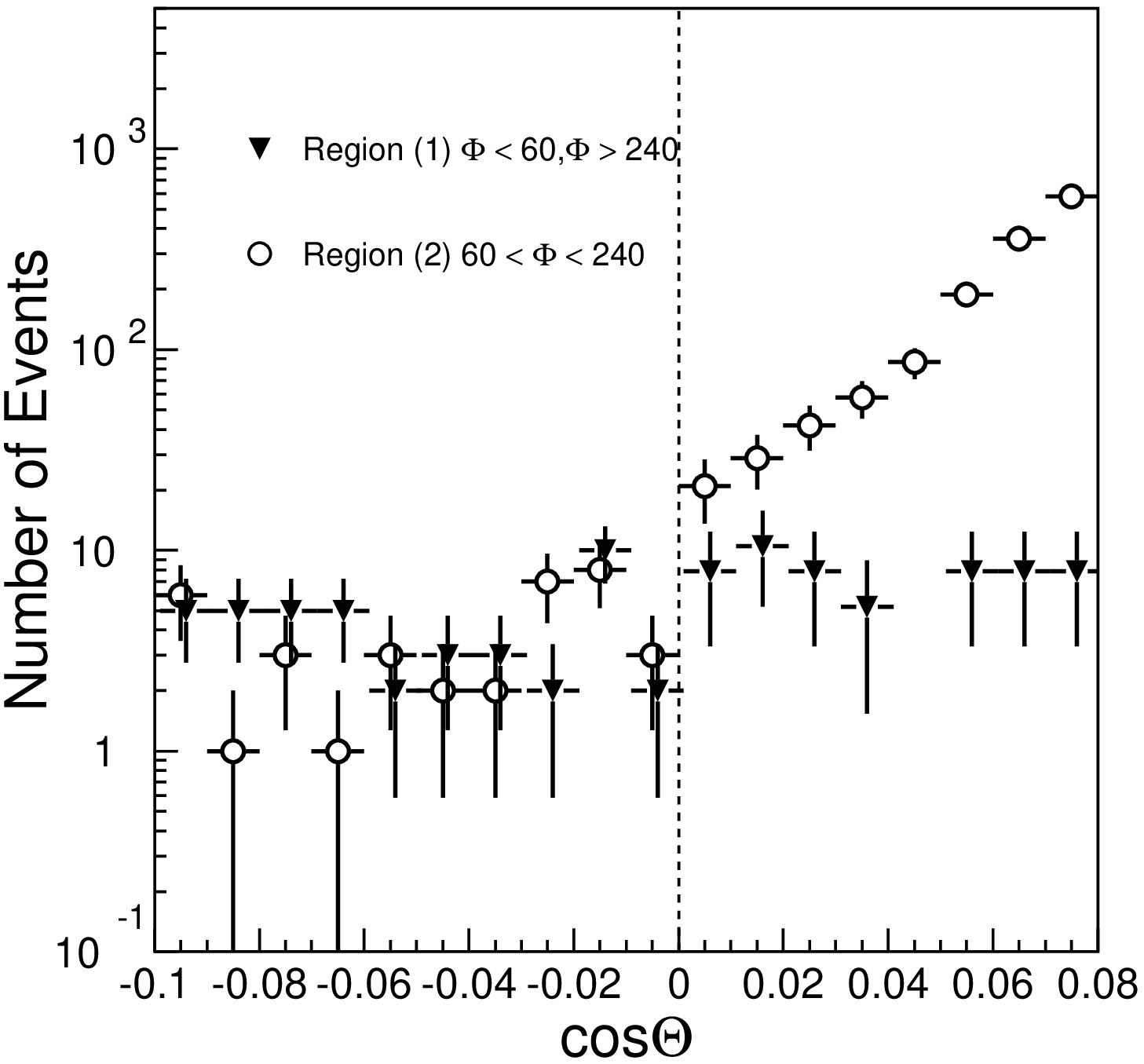}

\end{center}
  \vspace{-0.5pc}
  \caption{\label{nearhoriz} Zenith and azimuthal distribution of
    showering muons in the region of $-1.0 < cos\Theta < 0.08$.
Regions (1) and (2) in bottom panel  refer to thick and thin parts of the mountain respectively. }
\end{figure}

When we applied the showering muon algorithm to the upward through-going muon
sample from the 1646 days of data, we found a total of 318 showering events. The upward
muon data sample is contaminated near the horizon from downward going cosmic
ray muons which appear as upward going because of multiple Coulomb scattering
and the finite angular resolution of the fitters.  To estimate this 
background in
the showering muon sample, we use the same procedure as that applied for
stopping and through-going muons~\cite{fullpaper}. We apply the showering
algorithm and select a subsample of near horizon downward through-going 
muons
with $0 < \cos \Theta < 0.08$, where $\Theta$ is the zenith angle defined
with the vertical upward direction at $\cos \Theta = -1$. We then estimate
the background from the downward muon sample by extrapolation below the 
horizon~\cite{fullpaper}.  The zenith and azimuthal
distribution of near-horizontal showering muon events is shown in
Fig.~\ref{nearhoriz}. The expected background from showering muons is
estimated to be $9.15^{+13.0}_{-5.3}$, all in the bin from $0< \cos \Theta
<-0.1$. This background is subtracted from the upward showering muon
dataset for oscillation studies. For astrophysical studies, we assign 
a weight to this zenith angle bin  which is equal to the ratio of 
background-subtracted to
non-background subtracted events.
 
\begin{figure}
\begin{center}
    \includegraphics[height=17.0pc]{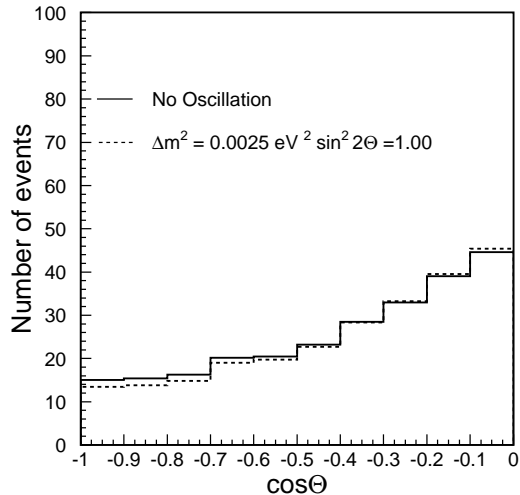}
\end{center}
  \vspace{-0.5pc}
  \caption{\label{zenithmc} The expected zenith angle distribution of 
    upward showering events with and without neutrino oscillations. 
Both histograms are normalized to the livetime of the dataset.}
\end{figure}

\section{OSCILLATION ANALYSIS}
Previous analyses have shown that the Super-K atmospheric neutrino dataset  is consistent 
with
neutrino oscillations with $\Delta m^2 \simeq 0.0025~\mbox{eV}^2$ and $\sin^2
2\theta = 1.0$~\cite{fullpaper,sklovere}. Given these neutrino oscillation
parameters, the oscillation probability is negligible for neutrinos with an
energy of 1 TeV at all path-lengths through the Earth. The zenith angle
distributions of upward showering muons without oscillation and with oscillated parameters 
obtained from Ref.~\cite{fullpaper} are shown in Fig.~\ref{zenithmc}. As expected, the difference between these distributions is negligible.
 We will demonstrate this by
comparing the zenith angle distribution of data and Monte Carlo.

The chi-square function used for comparison of data and Monte Carlo is
the same as that used for our published oscillation 
analysis~\cite{fullpaper}, which is based on the pull
method~\cite{Fogli}:

\begin{eqnarray}
\label{chi2def}
&\chi^2 &= \sum_{i=1}^{10} \left[\frac{N_i^{obs} - N_i^{exp}}{\sigma_i^{stat}}\right]^2 + 
\sum_{j=2}^{N_{sys}}\left[\frac{\epsilon_j}{\sigma_j^{sys}}\right]^2 \\
\label{oscexp}
&N_i^{exp} &= N_i^{osc}(1 + \sum_{j=1}^{N_{sys}}{f_i^j \epsilon_j})  
\end{eqnarray}
In Eqn.~\ref{chi2def}, $N_i^{obs}$ is the number of observed events in
$i^{th}$ bin, $N_i^{exp}$ is the expected number of events considering both
oscillation and systematic uncertainties,  $\sigma^i_{stat}$ combines
statistical uncertainties in the data and Monte Carlo simulation, $N_{sys}$
is the number of systematic errors used for the fit,
and $N_i^{osc}$ is the expected
number of events in this bin assuming $\nu_{\mu}$ to $\nu_{\tau}$ oscillation 
without considering the contribution from
systematic uncertainties. 
The factor $f^i_j$ represents the
fractional change in the predicted event rate in the $i^{\rm th}$ bin due to
a variation of the parameter $\epsilon_j$. The second sum in the $\chi^2$
definition collects the contributions from the systematic uncertainties in
the expected neutrino rates.

As we are doing oscillation analysis with only one data sample, and since
the absolute normalization is considered to be free as done in Ref.~\cite{fullpaper}, we only need to consider
those systematic error coefficients for which  the response is non-uniform in different zenith 
angle bins. The effect of all systematic  terms for which $f_i^j$ is same for all $i$ can be incorporated 
in the absolute normalization. Of all the systematic terms used in Ref.~\cite{fullpaper}, there are only
three  for which the response is different in various zenith angle bins. 
 These are shown in Table~\ref{chisquaretable}.

The atmospheric neutrino Monte Carlo does not include the attenuation of neutrino flux in the 
Earth. We did an estimate of the effect of neutrino absorption for the showering muons using the
models in Ref.~\cite{Gandhi} and the expected decrease in number of events is about 1.1\% . 
Therefore we do not incorporate it into the zenith angle prediction. Hence $N_{sys} = 4$ in Eqns.~\ref{chi2def} and
~\ref{oscexp}.

We vary these four~$\epsilon_j$ in order to minimize $\chi^2$ for
each choice of oscillation parameters $\sin^2 2\theta$ and $\Delta m^2$.
Among these, only three contribute to the $\chi^2$ because there are no constraints on the absolute
normalization. The minimum $\chi^2$ value, $\chi^2_{min} = 3.54 / 7 {\rm ~DOF}$, is 
located
at $(\sin^22\theta = 1.0,$ $\Delta m^2 = 1.05\times10^{-2} $~eV$^2$).  The
number of DOF is found by 10 terms in the $\chi^2$ sum plus three systematic
constraints in the $\chi^2$ sum minus four minimized parameters minus the 
two
physics parameters of $\sin^2 2\theta$ and $\Delta m^2$.  For null
oscillation, ($\sin^22\theta = 0$), we found a $\chi^2$ value of 6.21 for
9~DOF, where only the overall normalization is a free parameter. The best fit values
for the systematic uncertainties for null oscillations are indicated in 
Table~\ref{chisquaretable}. In Eqn.~\ref{oscexp}, the best-fit value of $\epsilon$ for absolute
normalization for the case of null oscillation is about 20.4\%.  The
minimum $\chi^2$ is not significantly different than the $\chi^2$ obtained
for null oscillation.

\begin{table*}[t]
\begin{center}
\caption {\label{chisquaretable}Summary of systematic errors and the best fit at null oscillation used for
upward  showering muons.}
\begin{tabular}{|c|c|c|c|}
\hline
No & Systematic uncertainty & $\sigma (\%)$  & Best-fit(\%) \\ \hline
1 & Absolute Normalization & Free & 20.4\\
2 & $K/\pi$ ratio & 20 & -6.75 \\
3 & Axial vector Mass ($M_A$) & 10 & 0.15 \\
4 & Multi-pion production (model-dependence) & 1 & -0.44 \\ \hline

\end{tabular}
\end{center}
\end{table*} 





\begin{figure}
\begin{center}
    \includegraphics[height=17.0pc]{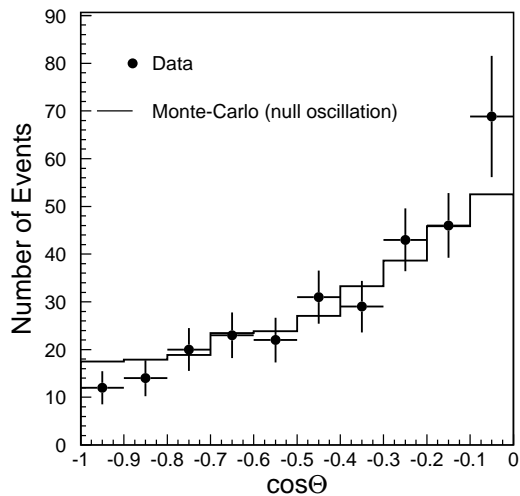}
\end{center}
  \vspace{-0.5pc}
  \caption{\label{zenith} Observed number of upward showering muons 
 as a function of zenith angle. The dots represent the observed events
with statistical error bars. 
The solid line shows the expectation at null oscillation evaluated using Eqn.~\ref{oscexp} 
and from best fit values in Table~\ref{chisquaretable}.}
\end{figure}

Thus, we satisfy an important consistency check of the neutrino oscillation
parameters determined in Ref.~\cite{fullpaper}, namely, that this dataset (with
mean neutrino energy of $\simeq$ 1 TeV) is consistent
with null oscillation.  
In future studies, this dataset could serve as an additional high energy bin, constraining 
high
$\Delta m^2$ solutions as well as certain non-standard oscillation scenarios~\cite{Nunokawa,Choubey}.  The zenith 
angle distributions of upward showering
muon data along with the expected distribution from null oscillations and using the best fit values
from Table~\ref{chisquaretable} are shown in Fig.~\ref{zenith}.

\section{ASTROPHYSICAL SEARCHES}
The high energy nature of this subset of the Super-K neutrino data provides
an advantage when searching for astrophysical neutrino point sources,
neutrinos from WIMP annihilation, and diffuse neutrino flux from the
galactic plane.  The main advantage in using this high energy sample for
astronomy is that the atmospheric neutrino background is very much reduced
due to the steeply falling atmospheric neutrino
spectrum~\cite{mannheim,hooper02,Stasto,Becker}.  Furthermore, the high momentum of the
incoming neutrino results in a muon which points more closely to the initial neutrino
direction at higher energies. The mean angular separation
between an upward showering muon and its parent neutrino is $2.1^{\circ}$
(assuming an atmospheric neutrino spectrum.) The corresponding numbers for
upward through-going muons and upward stopping muons are approximately $2.9^{\circ}$ and
$8.7^{\circ}$ respectively.  We also estimated the mean angular direction
between upward showering muons and their parent neutrinos for different
parent neutrino energy spectra. For cosmic ray spectra and $1/E^2$ spectra,
the mean angular separation is $1.8^{\circ}$ and $1.7^{\circ}$ respectively.
Therefore, we shall redo the searches for steady state point and diffuse
astrophysical sources from Ref.~\cite{sknuastro} with the showering muon
dataset.  There is no additional advantage in using this dataset for
transient astrophysical searches, since the timing window coincidence already
provides a strong cut to reduce the background.

\subsection{WIMP searches} 
We have performed searches for WIMP annihilations in the center of the Earth, Sun and Galactic Center 
using upward through-going muons~\cite{skwimp}. Here
we repeat the same search using upward showering muons. The cone size which contains
most of the WIMP signal is inversely proportional to the WIMP mass. 
Since only high-mass WIMPs produce upward showering muons, we perform these searches in cones only
up to $5^{\circ}$. Such a cone contains 90\% of the signal for a WIMP of mass 
1438 GeV from the Earth and 1000 GeV
from the Sun and Galactic Center~\cite{skwimp}. The observed data and expected background 
(evaluated in the same way as in Ref.~\cite{skwimp}) are shown in Table~\ref{wimptable}. 
Since, there is no statistically significant excess in any of the search cones, we do not see
any evidence for WIMP-induced upward showering muons in our dataset. 

\begin{figure}
\begin{center}
\includegraphics[height=17.0pc]{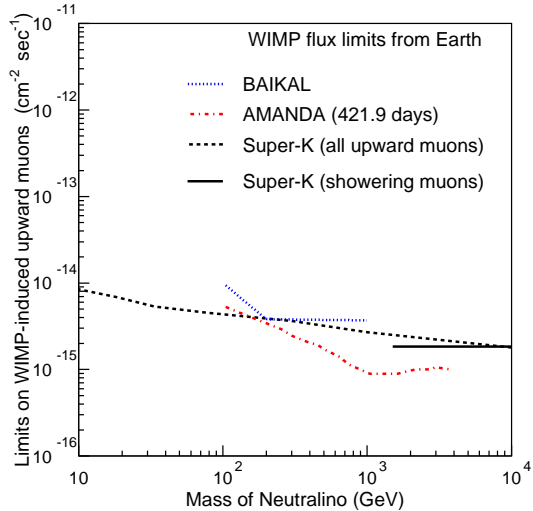}
\end{center}
\caption{\label{wimpearth}Super-K WIMP-induced flux limits of
upward showering muons and all upward muons from the Earth as a function of mass.}
\end{figure}

\begin{figure}
\begin{center}
\includegraphics[height=17.0pc]{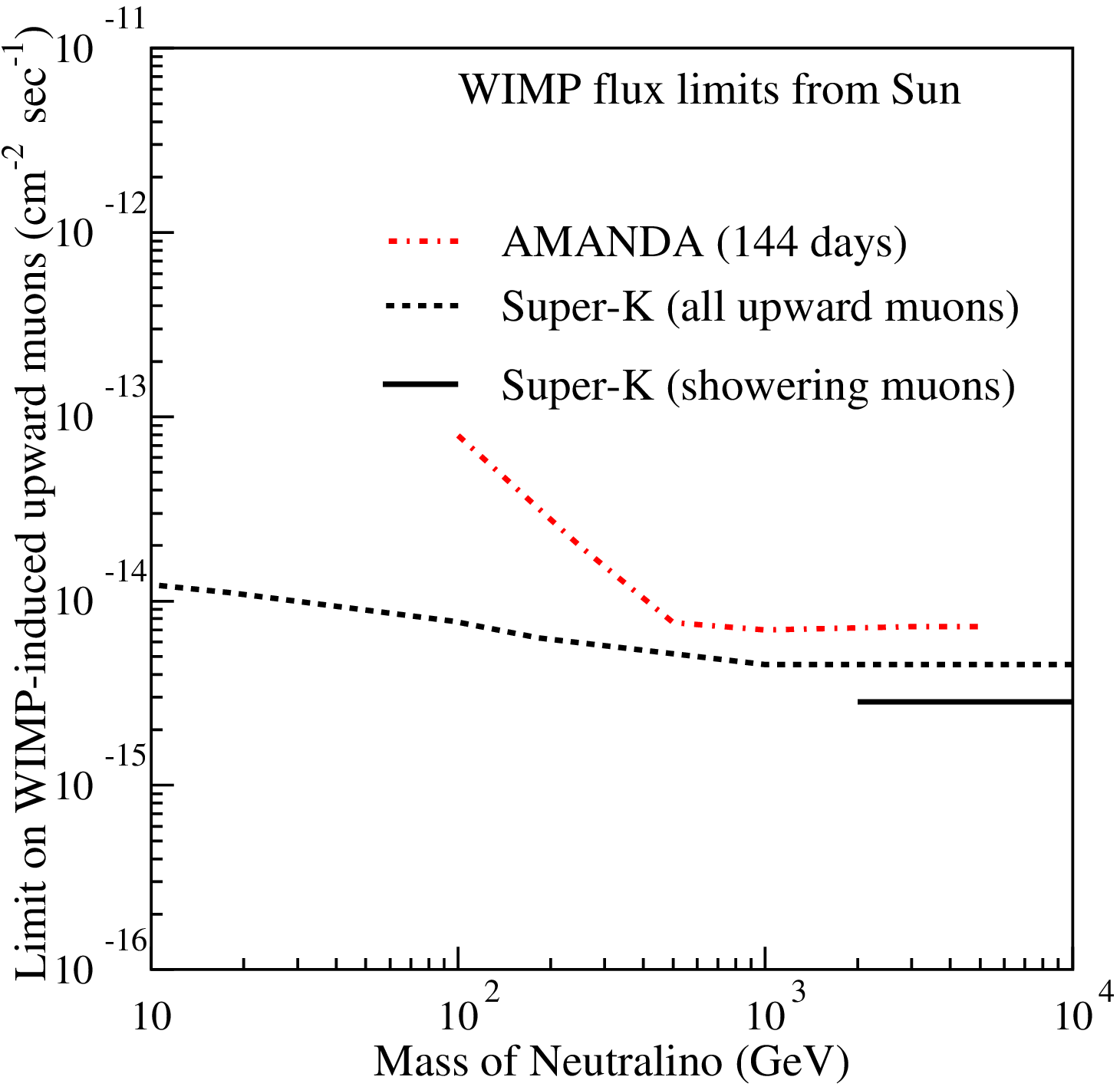}
\end{center}
\caption{\label{wimpsun}Super-K WIMP-induced flux limits of 
  upward showering muons and all upward muons from the Sun as a function of mass.}
\end{figure}

\begin{figure}
\begin{center}
\includegraphics[height=17.0pc]{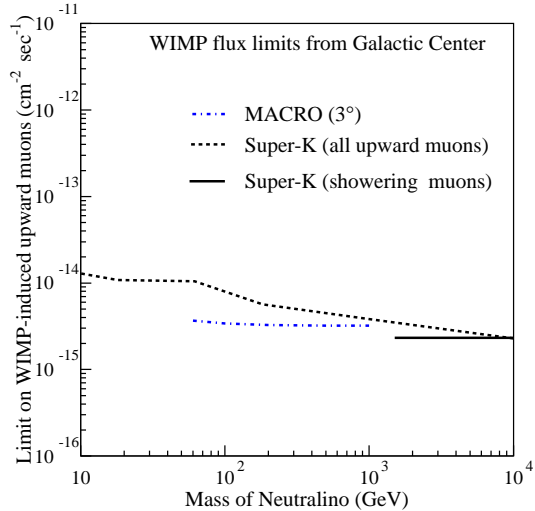}
\end{center}
\caption{\label{wimpgc}Super-K WIMP-induced flux limits of
  upward showering muons and all upward muons from the Galactic Center as a function of mass.}
\end{figure}

Because
of the reduced background from atmospheric neutrinos in this data sample, we
expect to obtain better flux limits with this sample, compared to those in
Ref.~\cite{skwimp}. These flux limits calculated using only upward showering
muons are plotted for the Earth, Sun and Galactic Center in
Fig.~\ref{wimpearth},~\ref{wimpsun}, and ~\ref{wimpgc}, respectively. The cutoff
WIMP mass used for the calculation of WIMP flux limits using only the showering muon 
dataset is 1500 GeV for the Earth and the Galactic Center, and 2000 GeV
for the Sun. This cutoff mass was calculated by determining the minimum mass for which
the neutrino energy spectrum from WIMP annihilation is peaked at 1 TeV. This spectrum was 
calculated by assuming the same branching ratio in all the  available annihilation channels 
for a given WIMP mass, and by using the analytic expressions for neutrino energy spectra calculated in 
Ref.~\cite{jk94}. The reason for the slightly higher
cutoff for the Sun is because, unlike the Earth~\cite{jkg95} and Galactic 
Center~\cite{gondolo99},
energetic neutrinos lose energy or get absorbed by neutral and charged
currents in the Sun~\cite{jk94,edsjo95,cirelli05,barger07,edsjo07}. To evaluate the WIMP 
flux limits with showering muons, we evaluated the showering efficiency  at different WIMP
masses. Note that we have also
extended the flux limits with all upward muons down to 10 GeV (from 18 GeV in
Ref.~\cite{skwimp}) by including the contributions of upward stopping muons.
We have also shown for comparison the corresponding limits from AMANDA-II and Baikal for
the Earth~\cite{amandaearth,baikalearth}, AMANDA-II limits for the Sun~\cite{amandasun}, and 
MACRO limits for the Galactic Center~\cite{macroapj}. The WIMP flux limits from all other
detectors can be found in Ref.~\cite{skwimp}. 

\begin{table*}[t]
\begin{center}
\caption {\label{wimptable}Observed and expected upward showering muons in
cones with half-angles $3^{\circ}$ and  $5^{\circ}$ around the
Earth, Sun  and Galactic Center.}
\begin{tabular}{|c|cc|cc|cc|}
\hline
\multicolumn{1}{|c|}{}  & \multicolumn{2}{|c|}{Earth} & \multicolumn{2}{|c|}{Sun} & \multicolumn{2}{|c|}{Galactic Center} \\
\hline
Cone  & Data & Background & Data & Background & Data & Background \\ \hline
$3^{\circ}$  & 0   & 0.1 &  0     & 0.2    & 0     & 0.4 \\  \hline
$5^{\circ}$  & 0   & 0.5 &  0     & 0.5    & 0     & 0.8 \\ \hline
\end{tabular}

\end{center}
\end{table*}

\subsection{Search for a signal from  potential point sources}
\begin{figure*}
\begin{center}
    \includegraphics[height=17.0pc]{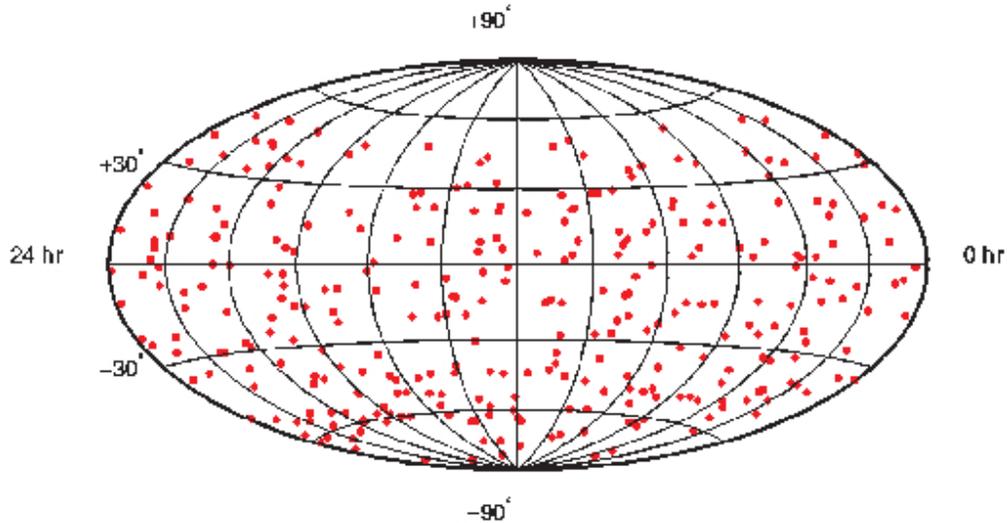}
\end{center}
  \vspace{-0.5pc}
  \caption{\label{skymap} Equatorial coordinate distribution of all upward
  showering muons.}
\end{figure*}

The equatorial coordinate distribution of upward showering muons is shown in
Fig.~\ref{skymap}. We look for signatures of neutrinos from 62 suspected point
sources. Fifty-two of these sources were analyzed previously in
Refs.~\cite{sknuastro,macroapj} \ and many of these satisfy some of the features of the
``beam dump model''~\cite{halzen95,Bednarek04}.  Most of these  sources are either supernova
remnants, pulsars, magnetars or different types of active galactic nuclei.
Some of these have been detected in TeV $\gamma$ rays~\cite{horan03}. We have
also considered some additional promising neutrino point sources following
the recent results from H.E.S.S $\gamma$--ray telescope \cite{Beacom,Kappes}.
  
In this search, we look for a statistically significant excess in a cone of
angular size $3^{\circ}$ around these sources. Such a cone size would contain 90 \%
of the signal for a $1/E^2$ neutrino spectra.
 The background was evaluated
using atmospheric neutrino Monte Carlo. Each Monte Carlo event was assigned a
time sampled from the upward showering muon distribution, in order to match
the livetime distribution of the observed events.  The observed data and
expected background are shown in Table~\ref{ptsourcetable}. As we can see, 
there is no statistically significant excess in any of these cones. 
For each source, we calculate the muon flux limit at 90\% c.l. using the method described
in Ref.~\cite{upperlimitref}.
To calculate the neutrino flux limits, we assumed a $E^{-2}$ neutrino energy
spectrum.
This spectrum arises in a number of astrophysical circumstances
such as  beam dumps and  shock acceleration.   
The calculation of neutrino flux limits from the showering muon flux limits is as follows.

The total flux of upward muons above an energy threshold ${E_{\mu}^{\rm min}}$ for a $1/E^2$ spectra is given by~\cite{halzen95}:
\begin{equation}
\label{muflux}
\Phi_{\mu} = A \int \limits_{E_{\mu}^{\rm min}}^{\infty} dE_{\nu} P_{\mu}\left(E_{\nu},\,E_{\mu}^{\rm min}\right) E_{\nu}^{-2}
\end{equation}

In Eqn.~\ref{muflux}, $A$ is a normalization constant for the differential 
neutrino flux, $P_{\mu}\left(E_{\nu},\, E_{\mu}^{\rm min}\right)$ is the probability 
that a neutrino with energy $E_{\nu}$  produces a muon with energy greater than $E_{\mu}^{\rm min}$. The 
total flux of upward showering muons $(\Phi_{\mu}^{showering})$ is then given by :

\begin{equation}
\label{showering}
\Phi_{\mu}^{showering} = k_{cone} k_{algo} \int \limits_0^{\infty} dE_{\mu}^{\rm min} \frac{d \Phi_{\mu} (\geq E_{\mu}^{\rm min})}{dE_{\mu}^{\rm 
min}} \epsilon(E_{\mu}^{\rm min}) 
\end{equation}

In Eqn.~\ref{showering}, $\epsilon(E_{\mu}^{\rm min})$ is the probability for a muon to 
undergo radiative energy losses  as a function of 
muon energy, $k_{algo}$ is the efficiency of our algorithm to detect a true showering 
muon using  the cuts in Sect.~\ref{showeringalgo}, and $k_{cone}$ is the 
fraction of the signal  which falls within the $3^{\circ}$ cone.
For each source, we solve for  $A$ by  substituting the obtained showering muon flux 
limits in the left-hand-side of Eqn.~\ref{showering}  
and using $\Phi_{\mu}$ as evaluated from Eqn.~\ref{muflux}.

\begin{figure*}
\begin{center}
\includegraphics[height=17.0pc]{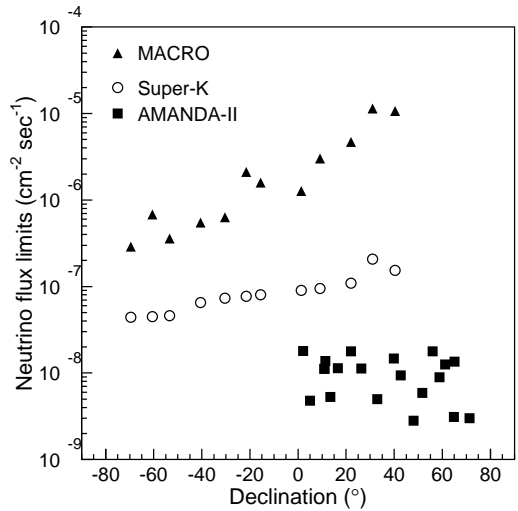}
\end{center}
\caption{\label{nuflux} Neutrino flux limits from Super-K, AMANDA-II~\cite{amandaptsource2} (for 
$E^{-2}$  neutrino spectra) and MACRO~\cite{macroapj}(for $E^{-2.1}$  neutrino spectra) as a 
function of declination for some selected sources. }  
\end{figure*}

To evaluate the above integrals,  we calculated the values of $P_{\mu}\left(E_{\nu},\, 
E_{\mu}^{\rm min}\right)$ using codes provided by M.~Reno~\cite{Gandhi} (2005,~private communication). 
More details on the assumptions used 
for this calculation can  be found in Ref.~\cite{skuheupmu}.  The 
efficiency $\epsilon(E_{\mu}^{\rm min})$ was calculated by first finding the fraction of 
muons from the mono-energetic Monte Carlo (used in Fig.~\ref{neutenergyloss}) 
with $dE/dX > $ 2.85 MeV/cm at 
different muon energies, after which we applied  a curve fit.  
The values of $k_{algo}$ and $k_{cone}$ were 
estimated for a $1/E^2$ spectra and each of them is equal to 0.9. Evaluating the integrals in 
Eqn.~\ref{showering}, the value of $A$ for each source is given by $A = 3.52 \times 
10^8~\Phi_{\mu}^{showering}~{\rm GeV^{-1}}$. Once $A$ is determined, the integrated neutrino 
flux limits were obtained by evaluating
$A  \int \limits_{E_{\nu}^{\rm min}}^{\infty} E^{-2} dE$, where  $E_{\nu}^{\rm min}$ is the minimum neutrino energy which would make  a 
showering muon and is chosen to be 10 GeV.
 This gives us 90\% c.l. neutrino flux limits for a given source 
from the corresponding 90\% c.l. showering muon flux limits.  These muon and neutrino flux limits for all the sources are shown in 
Table~\ref{ptsourcetable}. 
The corresponding neutrino
flux limits for some of these sources from MACRO and AMANDA-II  can be found in
Refs~\cite{macroapj} and Refs.~\cite{amandaptsource,amandaptsource2}. A plot showing a comparison 
of the neutrino flux limits from these detectors for some sample sources  as a function of declination 
can be found in Fig.~\ref{nuflux}.

\subsection{Search for unexpected point sources}
\begin{figure*}
\begin{center}
    \includegraphics[height=17.0pc]{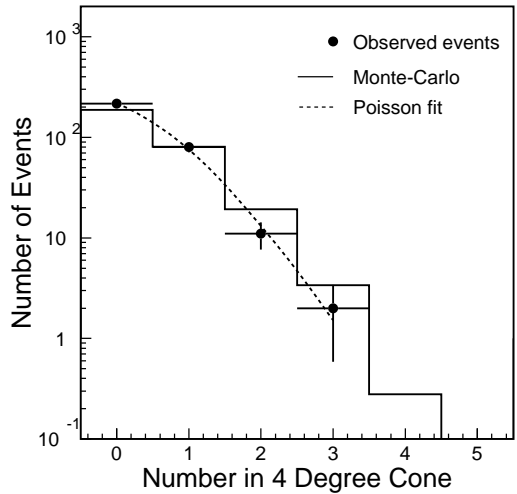}
\end{center}
  \vspace{-0.5pc}
  \caption{\label{unknown} Number of events observed 
 within a cone of half-angle $4^{\circ}$ of any showering muon(dots).
The solid line shows the expected atmospheric neutrino background. The dashed line shows a 
Poisson fit to the observed number of events with mean equal to 0.35 events}
\end{figure*}

In order to search for a signal from an unexpected point source (which may not have
any electromagnetic signature), we have looked
for an excess of events within a cone of half-angle $4^{\circ}$ around any
upward showering muon event. The background is estimated using the
atmospheric $\nu$ Monte Carlo in the same way as in Sect. 6.1.  The
distribution of the number of observed events with the $4^{\circ}$ cone is fit
well by a Poisson distribution with a mean of 0.35 events. The comparison of
data and Monte Carlo as well as results from the Poisson fit is shown in
Fig.~\ref{unknown}. Both methods show that the distribution of events
is consistent with the null hypothesis of having observed no signal from
unknown point sources.

\subsection{Diffuse Searches from the Galactic Plane}
Although all the expected cosmic ray-induced neutrinos observed at Super-K arise from 
interactions in the Earth's atmosphere, some neutrinos  could be produced 
from collisions of cosmic rays with hydrogen from the 
interstellar medium~\cite{Stecker79,Berezinsky93,Ingelman,Candia,Evoli}. 
At low energies the expected flux
from these neutrinos is negligible compared to atmospheric neutrinos.
According to some models~\cite{Ingelman}, the minimum energy at which the
atmospheric neutrino flux is equal to the flux of neutrinos produced from
cosmic ray interactions in the ISM is about 10 TeV.  Since most of the
hydrogen is concentrated in the galactic plane, a signature of such events
would be an excess from the galactic plane, which is characterized by galactic
latitude  of $0^{\circ}$. Searches for diffuse neutrinos from the galactic plane have been
done with Super-K using all upward through-going muons~\cite{sknuastro} and with AMANDA-II~\cite{AMANDAism}.
We now repeat this search with the showering muon dataset.

\begin{figure}
\begin{center}
\includegraphics[height=17.0pc]{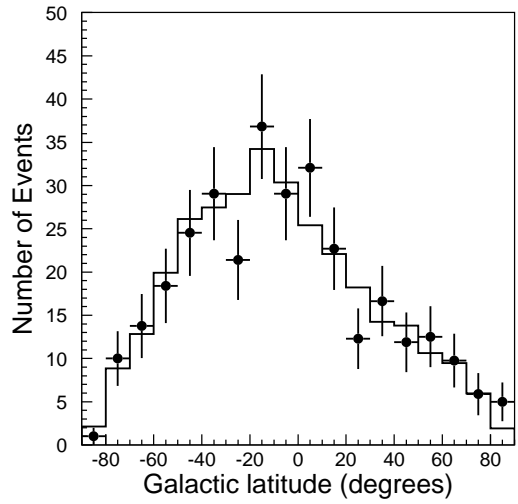}
\end{center}
\caption{\label{gc} Galactic latitude distribution of
  upward showering muons (dots) compared with Monte Carlo background from atmospheric 
neutrinos (solid line).} 
\end{figure}

The galactic coordinate distribution of upward
showering muons is shown in Fig.~\ref{gc}. As in Ref.~\cite{sknuastro}, we looked for a 
statistically 
significant excess within $\pm 10^{\circ}$ of the galactic plane. The observed signal events 
of 37 is consistent with the number of background events of 35, implying that there is no excess near galactic latitude of
$0^{\circ}$. Thus we do not see any evidence that any of the upward showering muon events are coming from interaction products of cosmic rays
with the interstellar medium.

\begin{deluxetable}{lcccc} 
\tablecaption{\label{ptsourcetable}Table of observed number of upward showering muons and expected 
background from atmospheric neutrinos in a cone of half-angle $3^{\circ}$ from selected point sources 
along with 90\% c.l. muon and neutrino flux limits for a $E^{-2}$ neutrino spectrum.}
\tablewidth{0pt}
\tablehead{
\textbf{Source}    &  \textbf{Data}     &  \textbf{Background} & \textbf{$\mu$ flux limits} & \textbf{$\nu$ flux limits} \\
\textbf{} & \textbf{} & \textbf{} &  \textbf{($10^{-15} cm^{-2} s^{-1}$)} & \textbf{($10^{-8} cm^{-2} s^{-1}$)}} 
\startdata
SMC X-1   & 0  &   0.3 & 1.3 & 4.4 \\
LMC X-2   & 0 & 0.3 & 1.3 & 4.4\\
SN 1987A  & 1 & 0.3 & 2.0 & 7.0 \\
LMC X-4  & 0 &  0.3 & 1.3 & 4.4 \\
GX301.2  &  0  & 0.6 & 1.3 & 4.5\\
Cen X-5  & 2 &   0.4  & 2.7 & 9.6\\
GX 304-1 & 1 & 0.4 & 2.0 & 7.0\\
Cen X-3 & 0  & 0.3 & 1.3 & 4.5\\
Cir X-1  &  1 & 0.6 & 1.9 & 6.8\\
2U 1637-53 &  0 & 0.3 & 1.3 & 4.6 \\
4U 1608-522  &  1 & 0.3 & 1.4 & 5.0\\
GX 339.4 &  0 & 0.3 & 1.6  & 5.6 \\
Vela   & 1 &  0.2 & 1.8  & 6.3\\
GX 346-7 &  0  & 0.2 & 1.7 & 6.0\\
AR X-1 &  0 &  0.2 & 1.7 & 6.1\\
SN 1006 &  1 & 0.4 & 2.8 & 9.8\\
Vela X-1 &  0 & 0.6  & 1.8  & 6.5 \\
2U 1700-37 &  0  &   0.3 & 1.9 & 6.7 \\ 
SGR X-4 & 1  & 0.2 & 2.0 & 7.3 \\ 
L10  & 0 &  0.4 & 2.1 & 7.2\\ 
GX 1+4 &   0 & 0.6 & 2.1 & 7.6\\ 
SN 1604 & 0 & 0.6  & 2.2 & 7.7\\ 
GX 9.9 & 0 &  0.4 & 2.3 & 8.0\\ 
Sco X-1 & 0 &  0.4 & 2.3 & 8.0\\ 
Aqr X-1  & 0 &  0.2 & 2.5 & 8.8\\ 
4U 336+01 & 1 & 0.3 & 4.0 & 14.1 \\ 
Aql  X-1 &  0 &  0.3 & 2.5 & 8.9\\ 
2U 1907+02 & 0 & 0.3 & 2.6 & 9.0\\ 
Ser X-1 &   0 & 0.2 & 2.6 & 9.2 \\ 
SS433  &  0 & 0.4 & 2.6 & 9.2 \\ 
2U 0613+09 & 0 & 0.1 & 2.7 & 9.5\\ 
Geminga  &  1 & 0.1 & 2.9 & 10.3 \\ 
Crab    & 0 & 0.3  & 3.1 & 10.9\\ 
2U 035+30  & 1 & 0.1 & 5.9 & 20.8\\ 
Cyg X-1  &  1 & 0.1 & 3.9 & 14.0\\
Her X-1  & 0 & 0.3 & 3.9 & 13.7\\
Mrk 421   & 0  & 0.2 & 4.2 & 14.8\\
Cyg X-2  & 0 &  0.1 & 4.2 & 14.7 \\
Mrk 501  & 0 &  0.2 & 4.4 & 15.4 \\
Cyg X-3 & 0 &  0.2 & 4.6 & 16.1 \\
Per X-1  & 0 & 0.2 & 4.7 & 16.4\\ 
SGR 1806 & 0 & 0.2 & 2.2 & 7.8 \\
SGR 1900 & 0 & 0.2 & 2.7 & 9.6\\
SGR 1627 & 1 & 0.3 & 2.9 & 10.1\\
SGR 1801 & 0 & 0.1 & 2.2 & 7.7\\
SGR 0525 & 1  & 0.3 & 2.0 & 7.0\\
LS 5039  & 1 & 0.4 & 3.5 & 12.4 \\
WR 20a   & 0 & 0.4 & 1.3 & 4.5\\
1ES 1959+650 & 0 & 0.2 & 2.7 & 9.4\\
B1509-58 & 1 & 0.4 & 2.0 & 7.0\\ 
B1706-44 & 0 & 0.3 & 1.7 & 6.1\\
B1823-13 & 2 & 0.4 & 5.0 & 17.5  \\
HESS J1303-631 & 1 & 0.4  & 2.0 & 6.9 \\
HESS J1514-591 & 2 & 0.4 & 2.7 & 9.6 \\
HESS J1614-518 & 1 & 0.3 & 1.5 & 5.2 \\
HESS J1632-478 & 0 & 0.3 & 1.6 & 5.7 \\
HESS J1702-420 & 1 & 0.1 & 1.8 & 6.4 \\
HESS J1745-303 & 0 & 0.2 & 2.1 & 7.2 \\
HESS J1804-216 & 0 & 0.3 & 2.2 & 7.7 \\
HESS J1825-137 & 1 & 0.4 & 3.6 & 12.6 \\
HESS J1834-087 & 1 & 0.3 & 3.8 & 13.4 \\
RX J0852-4622 & 0 & 0.3 & 1.7 & 6.0 \\
\enddata
\end{deluxetable}

\section{CONCLUSIONS}

From the sample of upward through-going muons in 1646 days of data
we have isolated a sample of 318 showering muon events which lose energy through
radiative processes. The mean parent neutrino energy of these events
is $\simeq$ 1 TeV  which make this subset the highest energy neutrinos
seen in Super-K. The fraction of these events is approximately one
fifth of upward through-going muon sample.  At these high energies, the 
neutrino oscillation probability (for oscillation parameters obtained from 
Ref.~\cite{fullpaper}) is negligible. We have verified that 
the zenith angle distribution of upward showering muons is consistent
with null oscillations. This shows that the highest neutrino energy dataset
(with mean parent neutrino energy of $\simeq$ 1 TeV, path-lengths of order
10000 km, and $\Delta m^2 \sim 0.0025$~eV$^2$) is consistent with null oscillation.    
We also performed various  searches for extra-terrestrial neutrino sources with this subset, 
such as from WIMP annihilations, suspected known and unknown point sources, as well as diffuse 
sources. We do not 
see any evidence for astrophysical neutrinos in the upward showering muon dataset.

\section{ACKNOWLEDGEMENTS}
We gratefully acknowledge the cooperation of the
 Kamioka Mining and Smelting Company. The Super-Kamiokande experiment
 has been built and operated from funding by the Japanese Ministry of
 Education, Science, Sports and Culture, the  United States Department  of 
Energy, and the U.S. National Science Foundation, with support for 
individual researchers from Research Corporation's Cottrell College Science Award.

\end{document}